# NEXUS/Physics: An interdisciplinary repurposing of physics for biologists

E. F. Redish[1], C. Bauer[2], K. L. Carleton[3], T. J. Cooke[4], M. Cooper[5], C. H. Crouch[6], B. W. Dreyfus[1], B. Geller[1], J. Giannini[1], J. S. Gouvea[1,7], M. W. Klymkowsky[8], W. Losert[1], K. Moore[1], J. Presson[9], V. Sawtelle[1], K. V. Thompson[9], C. Turpen[1,9], and R. K. P. Zia[10]

[1] Department of Physics, University of Maryland, College Park, MD 20742
[2] Department of Chemistry, University of New Hampshire, Durham, NH 03824
[3] Department of Biology, University of Maryland, College Park, MD 20742
[4] Department of Cell Biology & Molecular Genetics, University of Maryland, College Park, MD 20742
[5] Department of Chemistry, Michigan State University, East Lansing, MI 48824
[6] Department of Physics & Astronomy, Swarthmore College, Swarthmore, PA19081
[7] School of Education, University of California, Davis, CA 95616
[8] Department of Molecular, Cellular, & Developmental Biology, & CU Teach, University of Colorado, Boulder, CO 80309
[9] College of Computer, Mathematical & Natural Sciences, University of Maryland, College Park, MD 20742
[10]Department of Physics, Virginia Tech, Blacksburg, VA 24061 and Department of Physics and Astronomy, Iowa State university, Ames, Iowa 50011

***Abstract.*** In response to increasing calls for the reform of the undergraduate science curriculum for life science majors and pre-medical students (Bio2010, Scientific Foundations for Future Physicians, Vision & Change), an interdisciplinary team has created NEXUS/Physics: a repurposing of an introductory physics curriculum for the life sciences. The curriculum interacts strongly and supportively with introductory biology and chemistry courses taken by life sciences students, with the goal of helping students build general, multi-discipline scientific competencies. In order to do this, our two-semester NEXUS/Physics course sequence is positioned as a second year course so students will have had some exposure to basic concepts in biology and chemistry. NEXUS/Physics stresses interdisciplinary examples and the content differs markedly from traditional introductory physics to facilitate this. It extends the discussion of energy to include interatomic potentials and chemical reactions, the discussion of thermodynamics to include enthalpy and Gibbs free energy, and includes a serious discussion of random vs. coherent motion including diffusion. The development of instructional materials is coordinated with careful education research. Both the new content and the results of the research are described in a series of papers for which this paper serves as an overview and context.

## I. INTRODUCTION

Biology and the health-care sciences are playing an increasing role in STEM professions. There has been a strong growth in the number of biology and allied health undergraduate majors. They now represent one of the largest populations of students taking physics courses, nearly equaling or even surpassing the numbers of engineering students at some universities. In addition, the biology research community and medical schools have begun to call for improvements to the undergraduate education for life science majors, specifically emphasizing interdisciplinary coherence and the building of scientific skills and competencies.[1,2,3,4]

While there has been some interest in trying to design physics courses specifically for biology and medical students (hereafter, "life science students") for many years,[5,6,7] recently there has been a dramatic growth of interest in the physics community in an introductory course more explicitly designed for life science students – Introductory Physics for the Life Sciences (IPLS).[8]

Until recently, many physics departments have paid little attention to life science students, treating them as if they were mathematically challenged engineers and mixing them in with other populations, such as architects and computer scientists. Although many introductory physics texts claim to be "physics for the life sciences", a survey of these texts shows that mostly this has meant adding a few problems involving living organisms that are often simple rewrites of standard problems in a new context. This reflects the perception of many physicists that "physics is physics" and that essentially the same physics should be taught to all introductory students. Based on the work described in this and the associated papers cited in the references, we, along with many others,[8] have come to the conclusion that this perception needs to be changed.

In this paper, we describe a multi-year project, NEXUS/Physics, that brings together physics and biology education researchers, physicists, biologists, chemists, and biophysicists to consider these questions,

- What are the most important goals for a physics course for life science students?
- What can realistically be accomplished in a one-year course?
- How can physics be productively integrated with what life science students are studying in their biology and chemistry classes?

In this project, we are repurposing introductory physics so that it is more appropriate for life science students.

The *National Experiment in Undergraduate Science Education* – Project NEXUS – is based on a grant from the Howard Hughes Medical Institute and is complemented by an NSF collaborative TUES grant, *Creating a Common Thermodynamics* (CCΘ). NEXUS set up a four-university collaboration to develop materials for courses in physics (University of Maryland), chemistry (Purdue University), mathematics for biology classes (UMBC), and integrative case studies (University of Miami). (An overview of all four components is provided in K. Thompson et al.[9]) The CCΘ grant brings together disciplinary education specialists in physics, biology, and chemistry from six universities (University of Maryland, University of Colorado, University of New Hampshire, Swarthmore College, Michigan State University, and Virginia Tech).[10]

The project is inspired by the AAMC/HHMI report, *Scientific Foundations for Future Physicians* (SFFP), and the AAAS report, *Vision and Change* (V&C).[3,4] The goal of the physics component of the project (NEXUS/Physics) is to create a curriculum with supporting materials for life science students that:

- Supports the development of scientific reasoning skills and competencies as specified by the SFFP and V&C reports;
- Enhances students' perceptions of the value of scientific interdisciplinarity and their ability to use and build their knowledge across disciplinary boundaries;
- Relies on solid educational research on student learning.

In this paper, we describe the development and structure of NEXUS/Physics, the physics course created in conjunction with the NEXUS project. Many papers have been written and published on the research and development associated with the project. In what follows, we summarize briefly some of these results and put them in context. For more detail, refer to the original papers.

# II. THE PROCESS AND FRAMEWORK

## 1. Opening a dialog among stakeholders in biology, chemistry, & physics

To create the new course we:

- Built on previous projects of the University of Maryland (UMD) Physics & Biology Education Research Groups (PERG & BERG) that give us insight into what the students bring into the class and how we can focus on competency development;
- Held one year of extensive discussions among biologists, physicists, and chemists at Maryland and bi-weekly discussions among the members of the CCΘ project to decide what was needed;
- Carried out two years of materials development and testing of those materials in small classes;
- Carried out extensive ethnographic observations and education research through classroom videos, interviews, and analysis of artifacts. This research and previous research by the UMD PERG & BERG provided strong support for the new materials and were used to guide revisions;
- Established a close collaboration between the UMD-PERG and the UMD-Biophysics group, who are affiliated with all the departments involved in this interdisciplinary dialogue. The UMD Biophysics group provided numerous examples of biologically relevant physics that were developed into course activities, and led the development of the laboratories.

NEXUS/Physics was first delivered in 2011-12 and 2012-13 to small test classes. These were often taught using "flipped classroom" pedagogy; that is, students did preparatory work on their own time and then spent a significant fraction of class time engaged in active learning tasks. Students were assigned to read the written materials (a WikiBook) before class, write summaries, and ask a question. Class time was mostly used for discussion, clicker questions, problem solving, and some lecture. We videotaped all classes, enabling us to get a document how the students were responding to the material, both in interactions with the instructor and each other in the interactive lecture and with each other in discussions during recitation.

As a result of all that we learned from discussions with faculty and observations of students, we made the decision to shift the focus and overall purpose of the class. Since most of these students will *not* be taking future physics classes, the goal is not really to prepare them for later physics. We have therefore redefined our goal explicitly to be as follows:

> *The primary goals of NEXUS/Physics are to provide support for biology majors for difficult physics concepts that they will encounter in biology and chemistry classes, particularly those that cannot be studied in depth in those classes, and to help students build a sense of relevance. Our particular approach is to work with techniques suited to understanding science across the disciplines – using highly simplified models to build understanding, working carefully to build a sense of mechanism, developing coherences between things that might seem contradictory, etc.*
>
> *In addition, the course is specifically oriented toward competency development and interdisciplinarity. Many of the competencies stressed in biology, chemistry, and physics bridge the disciplines, but students need help to see how to connect them across disciplinary contexts.*

In the rest of the paper we provide a brief overview on how we implement these goals.

## 2. What we have learned from our research and conversations

We have learned some important lessons from our extensive discussions between biologists and physicists,[11,12] our studying students in biology and physics classes.[13,14,15]

- Biology faculty and physics faculty tend to have dramatically different views about the nature and structure of the knowledge that is appropriate to teach in introductory classes.
- Biology students tend to have difficulty and may even actively resist seeing the value of physics and mathematics in their learning of biology.

These are both tendencies, not absolutes. Still, the tendencies are common enough that we felt that we had to take them into account.

*Epistemological differences between physics and biology faculty*

The prime fact about biology is that it concerns living organisms and their interactions with each other and with the world they live in. A living organism is a complex system. One of the things that biologists learn quickly is that structure is often tightly tied to function, and that structure may be constrained not only by the laws of physics and chemistry, but by existing conditions arising from both evolutionary and random historical developments. Consequently, biology faculty prefer to discuss real specific systems – cells, organisms, ecosystems – and most examples require appreciating a great deal of realistic detail, even in introductory classes. There are few broad commonly agreed on principles in biology that have analytical power in specific cases

Another significant contrast to physics is that mathematical modeling is rarely introduced in introductory biology classes. A traditional 1000 page introductory biology textbook typically contains very few equations and those equations are rarely applied for mathematical reasoning or problem solving.[16] Even advanced undergraduate biology textbooks may contain few equations and little or no mathematical modeling.[17] This is beginning to change: more mathematically oriented texts are common in some upper level courses, and quantitative work is beginning to creep into some of the introductory texts as well.[18]

Physics takes a different approach to introductory classes. Mathematical modeling is considered fundamental and a 1000 page introductory physics textbook, even one that intends to emphasize "concept building", will typically have dozens or even hundreds of equations per chapter, thousands overall.[19] Fundamental principles, especially mathematically formulated ones, rule the roost and are used in wide varieties of situations in many different ways. Instead of emphasizing realistic situations, examples are emphasized that contain some essential character of real systems but that are mathematically tractable and help to develop fundamental insights. Such examples are often referred to as *toy models.* (The same phrase is used by professional physicists to characterize a similar approach in research contexts.) These highly idealized examples can then serve as starting points for creating more complex and realistic models.

Building and analyzing such models is typically seen by physics faculty to be one of the core skills of physicists. A major part of this skill is properly identifying what factors matter most in a complex system, incorporating those factors accurately in the model, and treating the factors that make the situation "more realistic" as corrections.

An implication is that many systems treated in physics are *abstracted* – general elements essential for the analysis at hand are stressed and objects are idealized. This is often carried to the point that seems ridiculous to someone not enculturated as a physicist. No one has ever seen a point mass, a perfectly rigid body, a massless string, or a spring that satisfies Hooke's law for any displacement. However, abstraction is valued by physicists because it permits one to identify factors that influence many phenomena without tying the result to a specific and conceptually limiting situation. Understanding such idealized systems helps one build and make sense of more realistic models of more complex systems."

These differing epistemological styles can lead to conflicts in trying to develop a physics class that speaks both to biology faculty and to biology students.

*Epistemological challenges to reaching biology students in a physics class*

Prior to the start of Project NEXUS, the UMD BERG & PERG studied student learning both in a biology class that attempted to include significant physics (Bio III: Organismal Biology)[12] and in a physics class that attempted to include applications to biology (Physics I & II: Algebra-Based Physics).[20] This was done with observations of students in class, extensive interviews, and pre-post attitude surveys exploring students' views on the roles of supporting classes for biology majors such as math, chemistry, and physics.[21]

These studies found that many students strongly resist both using physics and mathematical modeling in biology.

In a dramatic example,[15] one of our biology students, Ashlyn (all student names are pseudonyms), was studying the diffusion equation in her Organismal Biology class. One of Fick's Laws was presented that shows that in a time $t$ the distance that something diffuses is proportional to the square root of $t$. In a discussion with the interviewer on the role of math in biology, she resisted even this simple scaling argument.

> *I don't like to think of biology in terms of numbers and variables.... biology is supposed to be tangible, perceivable, and to put it in terms of letters and variables is just very unappealing to me.... Come time for the exam, obviously I'm going to look at those equations and figure them out and memorize them, but I just really don't like them.*
>
> *I think of it as it would happen in real life. Like if you had a thick membrane and tried to put something through it, the thicker it is, obviously the slower it's going to go through. But if you want me to think of it as "this is x and that's d and this is t", I can't do it.*

On the other hand, Ashlyn was successful in Physics I and II, earning A's, and later in the interview responded with considerable enthusiasm to a different scaling argument from her Org Bio class. A small model horse made of dowels and wooden blocks that stood quite nicely on the desk was scaled up in all dimensions by a factor of 2. The resulting larger horse's legs broke when it was placed on the desk as a result of the volume (and therefore the weight) going up faster than the cross-sectional area (and therefore the strength) of the legs.

> *The little one and the big one, I never actually fully understood why that was [before OrgBio]. I mean, I remember watching a Bill Nye episode about that, like they built a big model of an ant and it couldn't even stand. But, I mean, visually I knew that it doesn't work when you make little things big, but I never had anyone explain to me that there's a mathematical relationship between that, and that was really helpful to just my general understanding of the world. It was, like, mind-boggling.*

Ashlyn was representative of many students we have interviewed, both in our traditional and in the NEXUS/Physics classes. In many of our interviews, students were very accepting, and indeed enthusiastic, about physical and mathematical modeling when they perceived that applications had authentic relevance to the biology they were learning, but were dismissive and resistant when they perceived them as having little or no implications for biology.

We therefore conclude that incorporating *biologically authentic* applications is essential in winning life science students over to using physics and mathematical modeling. By *biologically authentic* applications, we mean those that use tools — such as concepts, equations, or physical tools — in ways and for purposes that reflect how the discipline of biology builds, organizes, and assesses knowledge about the world. We note that it is not only the perspective of the disciplinary expert that matters here; the *student's perception* of biological authenticity matters as well. When they perceive physics as valuable to their understanding of biology and chemistry, their engagement increases dramatically. Therefore, a primary goal of the changes we are making in NEXUS/Physics is for students to see this physics course as having biologically authentic value for them.

*Epistemological opportunities for making connections*

We, therefore, are paying particular attention in our redesign to creating opportunities that will help our students develop the view that physics helps them understand fundamental ideas in biology and chemistry, and that helps them see their biology and chemistry knowledge base as relevant to learning physics.[14]

## III. WHAT WE DECIDED: AN OVERVIEW OF OUR CHANGES

As a result of our conversations, we decided to make a number of changes from the traditional approach.

## 1. Rethinking the place in the curriculum

Many biology majors and pre-medical students do not take physics until their junior or even their senior year. This arises from two circumstances: (1) Physics is rarely a pre-requisite for any upper division class in their major.[22] (2) Their first two years are heavily loaded with other science and math classes. At Maryland, biology majors are required to take three semesters of introductory biology, four semesters of chemistry (two of General Chemistry and two of Organic Chemistry), and two semesters of calculus. Together, these two circumstances lead to physics as the course that often winds up being delayed.

As our goal is to integrate physics more coherently with courses in biology and chemistry, we are attempting to change this in two ways. First, we accept that physics will *not* be taken by biology majors in their first year. Second, we want upper division biology and chemistry instructors to see our physics as valuable enough to make it a prerequisite. To do this, we repositioned our class in the biology curriculum.

NEXUS/Physics is envisioned as a second year class for life science majors (perhaps starting in students' fourth semester). We require the following prerequisites:

- Two semesters of biology;
- One semester of general chemistry;
- Two semesters of calculus or math for biology majors.

We note that the initial courses on which we build are required of *all* biology majors at Maryland, whether they plan to specialize in cell/molecular biology, physiology, ecology & evolution, or medicine, and are widely seen both by the faculty and by our students in all of these areas as essential for their professional futures. While there are many additional physics topics that we could imagine as being particularly valuable for a one or more sub-populations, we cannot cover every potential application and therefore focus on topics that all biologists study.

For all of these prerequisites, we are not looking for a complete mastery of the materials in these classes so that we can build "on top of them". Rather, we are looking for sufficient exposure to the basic concepts so that they can be mentioned in a physics class. (Thus, it is NOT a prerequisite for faculty who intend to teach NEXUS/Physics that they have mastered the full content of the biology and chemistry prerequisites.)
We simply want students to have some sense of the following ideas.

From biology, we expect students to know basic ideas of anatomy and physiology (circulation of fluids, respiration, organs, bones), to understand that organisms are made up of cells, to have some sense of the structure of cells, and to have seen a discussion of some of the basic biochemical mechanisms such as DNA replication, the role of ATP, some ideas about cellular membranes, and a basic sense of evolution and how it works.

From chemistry, we expect students to know about atoms, molecules, and the types of chemical bonds. We also expect that they know that atoms and molecules have discrete energy levels. Finally, we expect students to have had an initial exposure to the ideas of entropy, enthalpy, and Gibbs free energy, and their relationship to spontaneity. As we discuss below, a potential value for our physics course is helping students make sense of these ideas, which are typically not taught in depth until much later in the chemistry curriculum.

In math, we expect them to know the basic ideas of calculus, including that a derivative is a rate of change and can be conceived of as a ratio of small quantities and therefore a slope on a graph; that an integral is the inverse of the derivative and can be conceived of as a product of a small quantity times the function value, summed up to give the area under a curve of a graph. We also make use of logarithms, exponentials, and probabilities and expect students to have some knowledge of these concepts and how to use them. These tools are used in a wide range of situations in upper division biology, ranging from the production of proteins in a cell to population dynamics.

Our focus is on building scientific skills and competencies in a way that creates a multi-disciplinary scientific approach.

## 2. Epistemological development: A focus on competency building

The NEXUS/Physics course starts from ten years of research and development done in the *Learning How to Learn Science* class (LHLS), an algebra-based physics class created by the UMD PERG through a series of NSF-supported projects.[20,23,24,25] This led to a course that focused on epistemological development, competencies, and general skills (*e.g.*, sense-making, seeking coherence, building intuitions) but maintained traditional introductory physics learning goals.

This course showed significant learning and attitude gains on standard measures,[20] but it neither made connections with what students learn in introductory biology and chemistry classes nor provided useful understandings for much in their upper division classes. The fact that we are now explicitly separating out life science majors allows us to better articulate with a their other science courses and address more specific epistemological goals.

Our more specific goals for this population include not just generalized competency development, but articulation with their classes in biology, chemistry, and mathematics, as identified in the SFFP and V&C reports.[3,4] We also want our students to:

- See the connectedness of the disciplines (not just applying physics to biology);
- Perceive relevance and utility of ideas from different disciplines;
- Identify problems and inconsistencies in what they know, and how to resolve them;
- Successfully reconcile diverse (sometimes apparently contradictory) disciplinary ideas.

Shifting the perspective and repositioning the class in the biology curriculum requires and enables a significant shift in the class content so that we can better integrate the physics with biology and chemistry.

## 3. Shifting the content

Our decisions to frame the course as integrating with and supporting biology and chemistry classes has led us to make some dramatic and potentially controversial changes in the focus of the physics course. Many physics faculty have constructed a coherent vision of a physics curriculum based primarily on their personal experience and on what they feel is needed to develop a coherent view of physics – one that they have in fact developed through more advanced physics classes.

But most of our life science students are not going to be taking more advanced physics classes. Most will major in biology. In response to this, physics faculty often respond with a (perhaps ex post facto) justification that the traditional course "exposes them to physics that they will see later in their careers" (such as angular momentum for understanding MRIs) and "shows them physics that is cool and inspirational" (such as universal gravitation and special relativity). While these are plausible goals for a physics class, they tend to produce a class that is "nice to have" but that is nowhere close to "important for a biology major".

When physicists look at the traditional course, they tend to see all of the content as obvious and natural. But when looked at in the context of biology, it becomes clear the traditional course includes many choices when other choices are possible – and better suited to biologists' needs. Some of our tacit assumptions as to what physics is essential seem better designed to mesh with a twentieth century engineering program. For example, in the past century, engineers dealt with construction using rigid materials such as metal and concrete, and with electrical circuits made of standard circuit elements. This leads to a physics course that emphasizes point masses, rigid bodies, heat engines, and Kirchoff's-law circuitry. Fluid flow considerations are restricted to pressure and basic rules of flow, as one would find in civil engineering classes.

In biological systems, however, fluids are everywhere and almost nothing is well approximated by a rigid body. Bones and the wood of a tree are pretty well approximated as rigid, but they are typically only components of more complex systems. The fundamental biology at the cellular or structural level rarely (not never!) is susceptible to treatment as a rigid body. Motion of objects within fluids is essential (since both the interior of cells and the environment in which most cells live is fluid) and the motion of electric charges in ionic solutions and across membranes is critical for many biological phenomena. The thermodynamics of chemical reactions is important while that of heat engines is mostly irrelevant.[26]

These perspectives have led us to redesign the content of the class. Some of the content changes (to be discussed in more detail in the next section) include:

- We restructure the range of phenomena to focus on dimensions from the nanoscale (atomic and molecular) to the macroscopic, with particular emphasis on the mesoscopic – (1 nm to 1 mm). Macroscopic phenomena at a scale much larger than any organism, such as rockets, universal gravitation, and planetary motion are eliminated;
- We reduce somewhat the discussion of forces and expand the discussion of energy, including discussing the energy in chemical reactions;
- We expand the discussion of fluids to include the Hagen-Poiseuille equation,[12] the motion of objects in a fluid (including "life at low Reynolds number"[27]) and electrical phenomena in ionic solutions;
- We include a careful treatment of random vs. coherent motion, and physical processes and concepts that arise from microscopic random motion, including Brownian motion, diffusion, entropic forces, and the Boltzmann factor; and
- We expand the traditional discussion of thermodynamics to include concepts students are exposed to in introductory biology and chemistry classes, including entropy, enthalpy, and Gibbs free energy.

In each of these cases, it is not our intent to provide students practice with the full mathematical machinery of these topics. Rather, because introductory physics is a place where toy models are accepted and explicitly used, our goal is to find simple but reasonably realistic examples where students can make better sense of concepts to which they have already been exposed in biology and chemistry by blending the conceptual with the basic mathematical ideas and representations.

Anyone who has taught a traditional introductory algebra-based physics course for life science students knows that the texts include far too much material to cover successfully in a single year. Everyone makes choices as to what to omit. As we are expanding the material to be covered considerably, some traditionally taught topics must be omitted. These were painful decisions, but we made them on the basis of two primary considerations:

- Does the topic have significant coherence with what the students are learning in their other classes (especially biology and chemistry), or does it seem "pasted on" – perhaps interesting, but irrelevant?
- Does the topic provide insight into a central and compelling biologically authentic problem?

Using these filters led us to cut back severely on or eliminate entirely the following topics: projectile motion, inclined planes, linear momentum, universal gravitation, statics, rotational dynamics, angular momentum, heat engines, alternating currents, magnetism, and relativity.

Cases can easily be made for the value of every one of these topics. For some populations of life science students, a case could be made for some of them being essential. We believe that for the broad range of biology majors and pre-meds, given the way that biology is increasingly coming to be taught (with an emphasis on the cellular scale and on biochemistry), that our choices are the most appropriate.

## IV. HOW WE GO ABOUT IT: A SUMMARY OF SOME SPECIFIC CHANGES

In this section, we discuss some of the primary changes that we have made in our IPLS course and briefly review some specifics of how these changes play out. There is not room in this short overview for a full discussion of the curriculum, how it is implemented in the classroom, and our data on the impact on student learning and attitudes. Instead, we provide brief descriptions of the main ways in which NEXUS/Physics differs from a traditional algebra-based physics class. Most of these differences arise from our attempt to better connect physics with what these students learn in their biology and chemistry classes. More complete discussions are presented in other papers that are referenced in the subsections below.

### 1. Respecting interdisciplinarity through including biological authenticity

To help students develop a sense that the biology they are learning is supported by a multi-disciplinary approach, we want tasks that are biologically authentic, but there is more to authenticity than just content. We have a set of explicit learning goals. We want students to

- Develop deeper levels of conceptual coherence that cross disciplinary boundaries;
- Develop a variegated set of scientific reasoning tools that may draw from different disciplines and can be applied in others;
- Develop adaptive expertise[28] in interdisciplinary thinking – the ability to draw upon knowledge and skills from multiple disciplines flexibly and apply them when appropriate;
- Develop positive attitudes about the value of the different disciplines.

For detailed discussions of these goals and examples of how they we use them as we construct specific tasks through a cycle of research and development, see Gouvea *et al.*[14] and Sawtelle *et al.*[29] The research and design cycle described there includes the following guidelines:

- Attend to differences in how disciplines represent and talk about big ideas (such as energy, entropy, work, etc.), but avoid collapsing into a singular language; instead, be explicit about and respectful of disciplinary differences;
- Identify similarities between reasoning strategies across disciplines;
- Engage students in seeking coherence among "things they know" across disciplines that appear to contradict each other (see for example, the discussions of chemical bonding and entropy in the next two sections);
- Engage students in re-evaluating their own and others' ideas; and
- Create space and time for students to grapple with and coordinate various representations, especially ones from different disciplines.

One example of disciplinary differences mentioned earlier is the treatment of simple models. Many of our students bring into our physics classes a reluctance to engage with what they see as oversimplified and unrealistic (toy) models. To explicitly engage these students in what they perceive to be a stark disciplinary difference, we are explicit about the role of modeling in physics,[30] explain why we begin with simplified models, and discuss (albeit briefly) the nature of the approximations we are making and the corrections that need to be made for more realistic discussions. We explicitly use *system schemas* taken from Modeling Instruction[31] that identify the objects in our model and their interactions.

An important circumstance that arises from these shifts is that students are now significant sources of knowledge in the class. Since a point of the course is to coordinate with what has been done in their biology and chemistry classes, and since the students are the best source of knowledge as to what they have learned there, there is a pedagogical shift. We do not expect physics instructors to be experts in all the biology and chemistry the students have learned. Rather, we expect that the students and the knowledge they bring in become important components in the learning environment. This requires that the instructor call for information and feedback from students and that the instructional environment be more responsive to the students.

## 2. Atoms, molecules, and chemical energy

One of the "big ideas" that appears in biology, chemistry, and physics classes is the concept of energy. A large part of the use of the concept of energy in introductory biology and chemistry classes involves the exchange of kinetic and potential energies with molecular rearrangement (reaction) energies and excitations. But as a result of the focus of traditional algebra-based physics classes on macroscopic phenomena, the topic of chemical reactions is never mentioned.[32] Because we have situated our class after biology and chemistry prerequisites, we can include it in our class.

A study of the chemistry education research literature,[33] our research with our students,[34,35,36,37] and extended discussions among biologists, chemists, and physicists convinced us that the topic was best treated by a coordination among the three disciplines.[38]

The chemistry education literature indicates that many students had difficulty in making sense of the source of energy in exothermic reactions. Specifically, they have difficulty reconciling the language used in biology classes of energy being "stored" in chemical bonds ("ATP is the energy currency of the cell") with the picture taught in chemistry of energy resulting from the fact that energy is required to break a bond and is released when a bond is formed.[39,40,41,42]

Our research with students led us to an understanding that there are (at least) three critical issues:

(1) Some students have difficulty making the conceptual connection between energy as discussed in standard physics classes at the macroscopic scale and the cellular and molecular scales as discussed in biology and chemistry classes;[34]
(2) Some students have difficulties seeing binding energies as negative and creating a coherent ontological picture of energy;[43]
(3) Discussions of chemical reactions taking place in physics and chemistry classes tend to make different tacit assumptions than discussions of the same reaction taking place in biology classes. Biology assumes that everything takes place in the context of an aqueous solution. Chemistry and (especially) physics tend to focus on isolated molecular reactions.

We address the first issue by integrating the atomic picture into our class throughout using macro (human scale ~1 m), micro (cellular scale ~$10^{-6}$ m), and nano (atomic and molecular scale ~$10^{-9}$ m) systems. Examples are drawn from each regime in every part of the class. Connections between molecular energies released in a reaction and the motion of human limbs are made in homework estimation problems and in group-work recitation activities.

We have begun to address the second issue by building a coherent development of the concept of energy throughout the class, starting with PhET simulations,[44] connecting to interatomic potentials, studying chemical reactions, discrete molecular bound states (phenomenologically), and spectroscopy. (Our chemical energy thread is discussed in detail in Dreyfus, et al.[37])

We address the third issue by raising the consciousness of the faculty to different disciplinary assumptions and making the contextual assumptions explicit in class discussions (often using the modeling structure and system schemas).[36]

## 3. Thermodynamics of entropy and free energy

Entropy and free energy often appear in first year biology courses and are essential components in chemistry beyond the first semester, but in these courses the treatment of the conceptual and mechanistic meaning of entropy is thin at best. Often, it makes its first appearance in the equation for the change in the Gibbs free energy ($G$) in terms of the change in enthalpy ($H$) and entropy ($S$)

$$\Delta G = \Delta H - T\Delta S$$

This equation is often introduced as a pragmatic tool for calculating whether a chemical process will spontaneously occur using tables of reference data. The discussion of entropy is often limited in these classes to the statement that it is a "measure of the disorder of the system."

Tensions arose for students in our course when their ideas about increasing entropy as representing disorder were brought into contact with their knowledge about the spontaneous formation of organized biological structures. If increasing entropy tends to make $\Delta G$ negative (and thus a process is spontaneous), how can organized biological systems form? Although chemistry texts have begun to move away from talking about entropy as "disorder" and towards describing it in terms of the dispersion of energy among accessible microstates, the language is still widespread and is being misinterpreted by students.

We address this tension through an approach that emphasizes the interplay of energy and entropy in determining spontaneity, one that involves a central role for *free energy* in an interdisciplinary thermodynamics curriculum. This approach draws on student resources from biology and chemistry in a particularly effective way. In conducting case study interviews over the past two years, perhaps no idea about thermodynamics was more strongly anchored and coordinated with other elements of our students' knowledge than the idea that spontaneity requires a negative change in the Gibbs free energy of a system. Most introductory physics texts don't mention free energy in association with spontaneity, and they certainly do not unpack the tension between energy and entropy (change in $\Delta H$ vs. change in $T\Delta S$) in leading to spontaneity.

A value for our course is in helping establish some basic conceptual knowledge about these concepts to support their further development in later chemistry and biology classes. These issues are discussed in more detail in Geller, et al.[45]

## 4. A statistical physics viewpoint

One perception grew strongly throughout our discussions as we explored the role of physics in modern molecular and cellular biology: the importance of statistical physics including a discussion of molecular processes that include non-equilibrium effects and fluctuations as argued by Klymkowsky, et al.[46,47] In support of the viewpoint of these authors, our perception of the value of these factors grew as a result of research interactions with biophysicists, a review of the research literature,[48,49,50] and conversations with leading statistical physicists who stressed the importance of non-equilibrium phenomena in biology.[51] Conversations with many biology and chemistry instructors confirmed that they perceived value in a combined conceptual and mathematical introduction to such subjects, despite the fact that, at the present time, most introductory biology textbooks (and animated lessons) ignore fluctuations and treat phenomena that are stochastic in nature as if they were purposeful and directed.[52]

As a result, in spite of the profound time constraints on a two-semester physics course, we carved out a place for a simple discussion of the tools of kinetic theory, building up to diffusion, the Boltzmann factor, and a careful discussion of wandering randomly through microstates as the basis of entropy and the second law. The theoretical treatments of these ideas in lecture and recitation are supported by and facilitate the study of random vs. coherent motion in the laboratories.

Although we have yet to complete an in-depth study of student learning on this topic, our preliminary evidence shows that students find the conceptual understanding of the mechanism of diffusion and microstates that they develop in physics to be very useful in their second year biology and chemistry classes.

## 5. Rethinking the laboratories

The laboratories for this class are built on the ideas developed in the LHLS studies. In those classes, laboratories were set up as *Scientific Community Labs* (SCL).[53] In these, students are not given a protocol as to how to do the lab step-by-step. Rather, they are given a single question. They work in groups of four to design an experiment to answer that question, build their apparatus, take data, and present their results to the other students in the lab as part of a discussion evaluating the different designs and approaches.[54]

The laboratories for NEXUS/Physics were transformed in a way that mimics the same kind of shifts that are taking place in the class while maintaining the overall structure of the SCL. The labs were designed to include significant content on physics relevant to cellular scales, from chemical interactions to random motion and charge screening in fluids. We also introduce the students to research-grade equipment (high power microscopes with video capture to study random motion) and modern analysis tools (Excel, ImageJ for video analysis) in contexts relevant to the life sciences.

Student responses to the laboratories have been very good, with almost all students agreeing that these labs are useful in preparing for their future careers. For more detail including descriptions of each lab and survey results, see Moore, et al.[55]

## 6. Publically Available Materials

In the three years of discussions, development, and testing, we have created a significant body of modular materials including:

- A WikiBook consisting of approximately 250 web pages,
- About 200 homework and group-work problems for recitations emphasizing biological content; these emphasize thinking and sense making and include no plug-and-chug (pick the equation and just put in numbers) problems,
- Collections of clicker problems to use in classes for generating discussion, and
- A set of laboratories.

We intend to continue generating more of these materials each time the class is delivered. All of these materials are presently available at the NEXUS/Physics website, *http://nexusphysics.umd.edu.* Readers and users of our materials are encouraged to send comments, criticisms, and suggestions to the senior author.[56]

In the long term, our intent is to make these materials available in whatever open IPLS distribution environment becomes available. (A number are currently under discussion by the IPLS community.) We are designing our environment so that it does not have to be considered as a fixed curriculum, but in a way that permits pieces of it to be used conveniently, that allows it to be combined as supplements to more traditional classes, and that will, hopefully, eventually turn into an environment that can grow organically and adapt to the changing needs of introductory physics for life scientists.

## V. CONCLUSIONS

As of this writing, we have primarily delivered the NEXUS/Physics class in our small development classes (2011-12 and 2012-13), with the first implementation in classes of several hundred taking place in the 2013-14 academic year. But from our analysis of our extensive data in our small classes (N~50), we can draw a few preliminary conclusions. For more details, especially some discussions of detailed case studies, see the cited papers.[29,36,37,43,45,55]

From our observations we conclude:

- Connecting the course material to biology and chemistry, through changes in content, placement within in the biology major sequence, and class discussions enables biology students to integrate physics into their overall scientific learning in a way that traditional physics classes do not.
- In this context, students bring sophisticated reasoning to physics from their other classes, identify inconsistencies in their knowledge that had previously gone unnoticed, and (with help) find ways to reconcile those inconsistencies.
- A class of the NEXUS/Physics type convinces many students of the value of physics in their scientific worldview.

An example of the second bullet comes from Gavin's comments in an interview that serve as a nice antidote to the discouraging comments quoted from Ashlyn earlier in the paper.

> *This class was very good about telling us about thermodynamics and entropy's role in the universe and why reactions proceed the way they do. And I think that diffusion was when everything kind of started to click. It is when we talked about how molecules go from higher concentration to lower concentration because they are bumping into each other and so these Newtonian interactions were able to move particles away from one another because the less they interact with each other, the more stable their environment really was... And...I felt like that is when things started to click. Oh, that's why molecules go from higher concentration to lower concentration.*

In our interviews, we saw many such responses from students.

The NEXUS/Physics course has gone through three years of discussion, development, and testing. We have learned a lot about how life science students engage in and learn from a physics class designed to address their needs. Understanding this learning process requires attending to not only where the students are coming from, but also to the pedagogical and curricular design features that support or hinder students' engagement and learning. In designing a course to bring multiple disciplinary perspectives into meaningful interaction, educational research played a central role. We did not "get it right" on the first try and it is only through careful observations, detailed analysis of student thinking, and multiple critical iterations that we are beginning to understand how to support students in this process. With appropriate supports integrated throughout such a course, we were amazed by what our students could accomplish. It is possible for biology majors to see learning physics as rewarding, engaging, and valuable to their future scientific pursuits.

## VI. ACKNOWLEDGEMENTS

This material is based upon work supported by the Howard Hughes Medical Institute NEXUS grant and the US National Science Foundation under Awards DUE 11-22818, DGE 07-50616, and partial support under DMR-12-44666. Any opinions, findings, and conclusions or recommendations expressed in this publication are those of the authors and do not necessarily reflect the views of HHMI or the National Science Foundation. We are grateful for conversations with numerous physicists, biologists, and chemists at the University of Maryland, and at the TRUSE, SABER, APS, and AAPT conferences.

[4] AAAS, *Vision and change in undergraduate biology education: A call to action*. (AAAS Press, 2011)

[5] A. P. French & E. F. Jossem. "Teaching physics for related sciences and professions," *Am. J. Phys.*, **44** (1976) 1149-1159. [http://ajp.aapt.org.proxy-um.researchport.umd.edu/resource/1/ajpias/v44/i12/p1149_s1]

[6] G. Benedek & F. Villars, *Physics, with Illustrative Examples from Medicine and Biology, 3 vols.* (preliminary edition, 1973; Springer, 2000).

[7] Philip Nelson, *Biological Physics: Energy, Information, Life* (Freeman, 2008); R. Phillips, J. Kondev, & J. Theriot, *Physical Biology of the Cell* (Garland Science, 2008).

[8] C. H. Crouch, R. Hilborn, S. A. Kane, T. McKay, M. Reeves, "Physics for Future Physicians and Life Scientists: a moment of opportunity", *APS News* (March 2010); note also a growing number of sessions at AAPT meetings and the development of specialized conferences (George Washington U., 2011; College Park, 2014).

[9] K. V. Thompson, J. A. Chmielewski, M. S. Gaines, C. A. Hrycyna, and W. R. LaCourse, Competency-based Reforms of the Undergraduate Biology Curriculum: Integrating the Physical and Biological Sciences, *Cell Biology Education - Life Science Education* **12** (June 3, 2013) 162-169. *doi:10.1187/cbe.12-09-0143*.

[10] NSF grant #DUE 11-22818, Collaborative Research: Creating a Common Thermodynamics.

[11] D. C. Meredith and E. F. Redish, Reinventing physics for life science majors, *Physics Today*, **66**:7 (2013) 38-43. *doi: 10.1063/PT.3.2046*

[12] E. F. Redish and T. J. Cooke, Learning Each Other's Ropes: Negotiating interdisciplinary authenticity, *Cell Biology Education - Life Science Education*, **12** (June 3, 2013) 175-186. *doi:10.1187/cbe.12-09-0147*.

[13] J. Watkins, J. E. Coffey, E. F. Redish, and T. J. Cooke, Disciplinary Authenticity: Enriching the reform of introductory physics courses for life science students, *Phys. Rev. ST Phys. Educ. Res.*, **8** (Apr 2012), 010112.

[14] J. S. Gouvea, V. Sawtelle, B.D. Geller, and C. Turpen, A Framework for Analyzing Interdisciplinary Tasks: Implications for Student Learning and Curricular Design, *Cell Biology Education - Life Science Education* **12** (June 3, 2013) 187-205. *doi:10.1187/cbe.12-08-0135*

[15] J. Watkins and A. Elby, Context dependence of students' views about the role of equations in understanding biology, *Cell Biology Education - Life Science Education* **12** (June 3, 2013) 274-286. *doi:10.1187/cbe.12-11-0185*

[16] S. Freeman, *Biological Sciences, 3rd Ed.* (Benjamin Cummings, 2008).

[17] B. Alberts *et al*., *Molecular Biology of the Cell, 4th Ed.* (Garland Science: Taylor & Francis, 2002).

[18] J. Feser, H. Vasaly, & J. Herrera, On the edge of mathematics and biology integration: Improving quantitative skills in undergraduate biology education, *CBE-LSE* **12**:2 (Summer 2012) 124-128;

[19] K. Cummings, P. Laws, E. Redish, & P. Cooney, *Understanding Physics* (John Wiley & Sons, Inc., 2003).

[20] E. F. Redish and D. Hammer, Reinventing College Physics for Biologists: Explicating an Epistemological Curriculum , *Am. J. Phys.*, **77**, 629-642 (2009). [supplementary appendix]

[21] K. Hall, *Examining the effect of students' classroom expectations on undergraduate biology course reform,* PhD dissertation, U. of Maryland (2013) [http://www.physics.umd.edu/perg/dissertations/Hall/]

[22] Except occasionally neurophysiology.

[23] L. Lising & A. Elby, "The impact of epistemology on learning: A case study from introductory physics," *American Journal of Physics*, **73** (2005) 372-382. [http://ajp.aapt.org/resource/1/ajpias/v73/i4/p372_s1]

[24] T. I. Smith and M. C. Wittmann, "Comparing three methods for teaching Newton's third law," *Phys. Rev. ST Phys. Educ. Res.* **3** (2007) 020105.

[25] A. Elby, *et al*., *Open-Source Tutorials in Physics Sensemaking* [http://umdperg.pbworks.com/w/page/10511218/Open%20Source%20Tutorials]

[26] One might argue that heat engines are an appropriate place to learn the basic concepts of thermodynamics. But these ideas do not find good traction with biology students and lead to the delay of more important concepts, such as Gibbs free energy and chemical thermodynamics, to be carefully dealt with in later courses such as Physical Chemistry (that many biology students do not take).

[27] E. M. Purcell, "Life at low Reynold's number," *Am. J. Phys.* **45**:1 (1977) 3-11.

[28] G. Hatano and K. Inagaki, "Two courses of expertise," in *Child development and education in Japan*, eds. H. Stevenson, H. Azuma, and K. Hakuta, (W. H. Freeman and Co. 1986) 262–272; J. W. Pellegrino, "Rethinking and redesigning curriculum, instruction and assessment: What contemporary research and theory suggests," http://www.skillscommission.org/pdf/commissioned_papers/Rethinking%20and%20Redesigning.pdf (2006, accessed February 15, 2008).

[29] V. Sawtelle, C. Turpen, B. Dreyfus, B. Geller, and J. Svoboda Gouvea, "Investigating student outcomes to refine interdisciplinary learning goals," submitted for publication (2013); C. Turpen & V. Sawtelle, "Iteratively designing an IPSL course to support building disciplinary connections," invited talk, AAPT National Meeting, Portland OR (July 15, 2013).

[30] D. Hestenes, "Toward a modeling theory of physics instruction," *Am. J. Phys.* **55**:5 (1987) 440-454.

[31] L. Turner, "System Schemas," *The Physics Teacher*, **41**:7 (2003) 404. doi:10.1119/1.1616480.

[32] This is not quite true. If you search the indices of many algebra-based physics texts for the life sciences, in

some you will find a single reference to chemical energy or reactions: a statement that it exists.

[33] B. Dreyfus, B. Geller, D. Meltzer, and V. Sawtelle, "Resource letter: Teaching thermodynamics and statistical mechanics in physics, chemistry, and biology," submitted for publication (2013).

[34] B. Dreyfus, E. Redish, and J. Watkins, Student Views of Macroscopic and Microscopic Energy in Physics and Biology, B *Proceedings of the Physics Education Research Conference, Omaha, NE, August 2011, AIP Conf. Proc.* **1413,** 179-182 (2012).

[35] B. W. Dreyfus, B. D. Geller, V. Sawtelle, J. Svoboda, C. Turpen, and E. F. Redish, Students' Interdisciplinary Reasoning about 'High Energy Bonds' and ATP , in *Proceedings of the Physics Education Research Conference, Philadelphia, PA, July 2012, AIP Conf. Proc.* **1513** (2013) 122-125.

[36] B. W. Dreyfus, V. Sawtelle, C. Turpen, and E. F. Redish, A Vision of Interdisciplinary Education: Students' reasoning about 'high-energy bonds' and ATP, submitted for publication (2013).

[37] B. Dreyfus, B. Geller, J. Gouvea, V. Sawtelle, C. Turpen, and E, Redish, "Chemical energy in an introductory physics course," *Am. J. Phys.,* in press (2014).

[38] M. M. Cooper and M. W. Klymkowsky, The Trouble with Chemical Energy: Why Understanding Bond Energies Requires an Interdisciplinary Systems Approach, *CBE Life Sci Educ* **12** (June 3, 2013) 306-312; *doi:10.1187/cbe.12-10-0170*

[39] S. Novick, "No energy storage in chemical bonds," *J. Biol. Educ.* **10** (1976) 116-118.

[40] H.K. Boo, "Students' Understandings of Chemical Bonds and the Energetics of Chemical Reactions," *J. Res. Sci. Teach.* **35** (1998) 569-581.

[41] M.A. Teichert and A.M. Stacy, "Promoting understanding of chemical bonding and spontaneity through student explanation and integration of ideas," *J. Res. Sci. Teach.* **39** (2002) 464-496.

[42] R.D. Storey, "Textbook errors & misconceptions in biology: Cell energetics," *Am. Biol. Teach.* **54** (1992) 161-166.

[43] B.W. Dreyfus, B.D. Geller, J. Gouvea, V. Sawtelle, C. Turpen, and E.F. Redish (2013), Negative gy: Why interdisciplinary physics requires multiple ontologies, *submitted to 2013 PERC Proceedings*; B. W. Dreyfus, B. D. Geller, J. S. Gouvea, V. Sawtelle, C. Turpen, and E. F. Redish, "Ontological Metaphors for Negative Energy in an Interdisciplinary Context," submitted for publication (2013).

[44] C.E. Wieman, W.K. Adams, K.K. Perkins, PhET: Simulations That Enhance Learning , *Science*, **322** (2008) 682-683.

[45] B. D. Geller, B. W. Dreyfus, J. S. Gouvea, V. Sawtelle, C. Turpen, & E. F. Redish, "Entropy and spontaneity in an introductory physics course for life science students," submitted for publication (2013); B.D. Geller, B.W. Dreyfus, V. Sawtelle, J. Svoboda, C. Turpen, and E. F. Redish (2012), "Students' reasoning about interdisciplinarity," *Proceedings of the 2012 Physics Education Research Conference, AIP Conf. Proc.* **1513,** (2013) 146-149; B.D. Geller, B.W. Dreyfus, V. Sawtelle, J. Svoboda, C. Turpen, and E.F. Redish "Like dissolves like: Unpacking student reasoning about thermodynamic heuristics," submitted for publication (2013).

[46] K. Garvin-Doxas, M. Klymkowsky, and S. Elrod, "Building, using, and maximizing the impact of concept inventories in the biological sciences: report on a National Science Foundation sponsored conference on the construction of concept inventories in the biological sciences," *CBE-Life Sciences Education*, **6** (2007) 277-282.

[47] K. Garvin-Doxas and M.W. Klymkowsky, "Understanding Randomness and Its Impact on Student Learning: Lessons Learned from Building the Biology Concept Inventory (BCI)," *CBE Life Science Education* **7**, 227 (2008); doi:10.1187/cbe.07-08-0063

[48] R. Phillips, J. Kondev, & J. Theriot, *Physical Biology of the Cell* (Garland Science, 2008).

[49] N Maheshri & EF O'Shea, Living with noisy genes: how cells function reliably with inherent variability in gene expression.Annual review of biophysics and biomolecular structure, 2007;36:413-34.

[50] V. B. Shahrezaei & B Swain, The stochastic nature of biochemical networks *Current opinion in biotechnology* 19(4): 369-74. Epub 2008 (Aug 2008).

[51] Thanks to Michael Fischer and Chris Jarzynski for highly useful conversations.

[52] A lovely antidote to these molecular anthropomorphizations are the animations being created by the Theoretical and Computational Biophysics Group at the University of Illinois Urbana-Champaign. [http://www.ks.uiuc.edu]

[53] http://umdperg.pbworks.com/w/page/10511229/Scientific%20Community%20Labs

[54] Rebecca Lippmann Kung, "Teaching the concepts of measurement: An example of a concept-based laboratory course," *Am. J. Phys*. **73** (2005) 771-777; R. F. Lippmann, *Students' Understanding of Measurement and Uncertainty in the Physics Laboratory: Social construction, underlying concepts, and quantitative analysis,* Ph.D. dissertation, U. of Maryland (2003) [http://www.physics.umd.edu/perg/dissertations/Lippmann/]

[55] K. Moore, J. Giannini, & W. Losert, "Toward better physics labs for future biologists," *Am. J. Phys.*, in press (2014); K. Moore, J. Giannini, B. Geller, W. Losert, "Physics for Biologists: A Laboratory Curriculum Project," contributed poster, AAPT National Meeting, Portland, OR (July 2013).

[56] Redish@UMd.edu. In the near term, we plan to create a site for instructors where comments and conversations can be carried out on the site itself.